\newif\ifCameraReady
\CameraReadyfalse

\documentclass[conference]{IEEEtran}
\IEEEoverridecommandlockouts
\usepackage{cite}
\usepackage{amsmath,amssymb,amsfonts}
\usepackage{algorithmic}
\usepackage{graphicx}
\usepackage{textcomp}
\usepackage{xcolor}
\def\BibTeX{{\rm B\kern-.05em{\sc i\kern-.025em b}\kern-.08em
    T\kern-.1667em\lower.7ex\hbox{E}\kern-.125emX}}
\begin{document}

\title{Identifying Impacts of Protocol and Internet Development on the Bitcoin Network\\
}

\author{\IEEEauthorblockN{Ryunosuke Nagayama, Ryohei Banno$^{\dagger}$, and Kazuyuki Shudo}
\IEEEauthorblockA{\textit{Tokyo Institute of Technology} \\
Tokyo, Japan \\
nagayama.r.ac@m.titech.ac.jp}
}

\ifCameraReady
\IEEEpubid{978-1-7281-8086-1/20/\$31.00 \copyright2020 IEEE}
\fi

\maketitle
\setlength\floatsep{-3pt}
\begin{abstract}
Improving transaction throughput is an important challenge for Bitcoin.
However, shortening the block generation interval or increasing the block size to improve throughput makes it sharing blocks within the network slower and increases the number of orphan blocks.
Consequently, the security of the blockchain is sacrificed. 
To mitigate this, it is necessary to reduce the block propagation delay.
Because of the contribution of new Bitcoin protocols and the improvements of the Internet, the block propagation delay in the Bitcoin network has been shortened in recent years.
In this study, we identify impacts of compact block relay---an up-to-date Bitcoin protocol---and Internet improvement on the block propagation delay and fork rate in the Bitcoin network from 2015 to 2019.
Existing measurement studies could not identify them but our simulation enables it.
The experimental results reveal that compact block relay contributes to shortening the block propagation delay more than Internet improvements.
The block propagation delay is reduced by 64.5\% for the 50th percentile and 63.7\% for the 90th percentile due to Internet improvements, and by 90.1\% for the 50th percentile and by 87.6\% for the 90th percentile due to compact block relay.
\end{abstract}


\renewcommand{\thefootnote}{\fnsymbol{footnote}}
\footnote[0]{$\dagger$ Current affiliation as of June 2020: Kogakuin University}
\renewcommand{\thefootnote}{\arabic{footnote}}

\section{Introduction}
Blockchain is a distributed system with Byzantine fault tolerance.
Because blockchain can manage a distributed ledger without a centralized system and has difficulty for tampering with past data, blockchain is used as the core technology for cryptocurrencies.
However, Proof-of-Work (PoW), an algorithm used in many blockchains such as Bitcoin, has the limitation that it can process only a small number of transactions at a given time.

One solution to this problem is to shorten the block generation interval; however, this sacrifices blockchain security\cite{b1}.
If the block generation interval is shortened, it becomes difficult to share the block in the network, and inconsistency in the blockchain arises.
Therefore, it is essential to reduce the block propagation delay to share a block in a short interval.

According to a monitoring web site\cite{b2}, the Bitcoin block propagation delay has been shortened in recent years.
The block propagation delay has decreased from more than 5 seconds in 2015 (until 50\% of peers received a block) to less than 1 second in 2019.
Further, the 90th percentile decreased from more than 15 seconds in 2015 to approximately 2 seconds in 2019. 

There are several reasons for the decrease in the block propagation delay\cite{b3}.
The first reason is the use of relay networks, such as Falcon\cite{b4} and FIBRE\cite{b5}, which are structured to efficiently propagate blocks and transactions.
Second, development of the Bitcoin protocol, such as compact block relay (CBR)\cite{b8}, reduce the block size and enable high-speed block transmission by sending only the transaction~ID, not the entire transaction.
Furthermore, improvements to the Internet have reduced network latency between peers, increased bandwidth, and reduced data communication time\cite{b12,b13}.

Otsuki et al.\cite{b6} analyzed the impact of relay networks on the block propagation delay and fork rate by simulating a relay network on the Bitcoin network using the blockchain network simulator SimBlock\cite{b7,b15}.
However, there has been no quantitative analysis of the impacts of protocol development such as CBR and Internet improvements on reducing the block propagation delay.
Existing measurement studies could not identify impacts of them because measured bandwidths and latencies reflect them together.
In this study, we identify impacts of compact block relay and Internet improvements on the block propagation delay and fork rate in 2015 and 2019 by simulating them.

The remainder of this paper is organized as follows. In Section II, we provide an overview of Bitcoin, while in Section III we describe a simulator, a procedure for calculating network parameters, and CBR models. In Section IV, we discuss our experimental results, and in Section V, we provide conclusions and present ideas for future work.

\ifCameraReady
\IEEEpubidadjcol
\fi

\section{Bitcoin Network}
\subsection{Block generation}
Bitcoin uses a data structure called a block to record transactions.
Nodes store transactions broadcast to the network in a memory pool and generate blocks from them.
The generated blocks are broadcast and the receiving node validates and adds them to the blockchain.
In addition to the transaction, a block contains the hash value of the previous block, which makes the blockchain a history of successive transactions.
A block also contains a value called a nonce. In the PoW algorithm adopted by Bitcoin, generating a block is equivalent to identifying a block whose total hash value falls below a certain threshold while changing the nonce. This process is called mining.
Currently, the block generation interval is adjusted to approximately 10 min according to the difficulty that determines the threshold.

However, the blockchain may be forked by generating multiple blocks from the same parent block.
In this case, receiving nodes select a block with the largest total difficulty of all blocks up to that block as the head block.
A blockchain that comprises ancestor blocks of the head block is considered to be a legitimate blockchain.

If malicious nodes attempt to change an approved transaction in a past block, nodes must generate blocks until the total difficulty exceeds that of the current legitimate blockchain.
Therefore, the longer the chain is between the block containing the transaction to be changed and the current head block, the more difficult it is to change the transaction.

\subsection{Block propagation}
When a node generates a block by mining, the node sends the generated block to neighboring nodes.
There are currently two main protocols for block propagation in Bitcoin: the legacy protocol developed for the first implementation of Bitcoin and CBR. 

In the legacy protocol, nodes propagate all transactions contained in a block, which requires significant network resources, typically close to 1 MB ($10^6$ bytes) per block\cite{b9}.
In contrast, CBR is a protocol for reducing the amount of bandwidth used to propagate new blocks to nodes.
\begin{figure}[t]
    \centerline{\includegraphics[width=\linewidth]{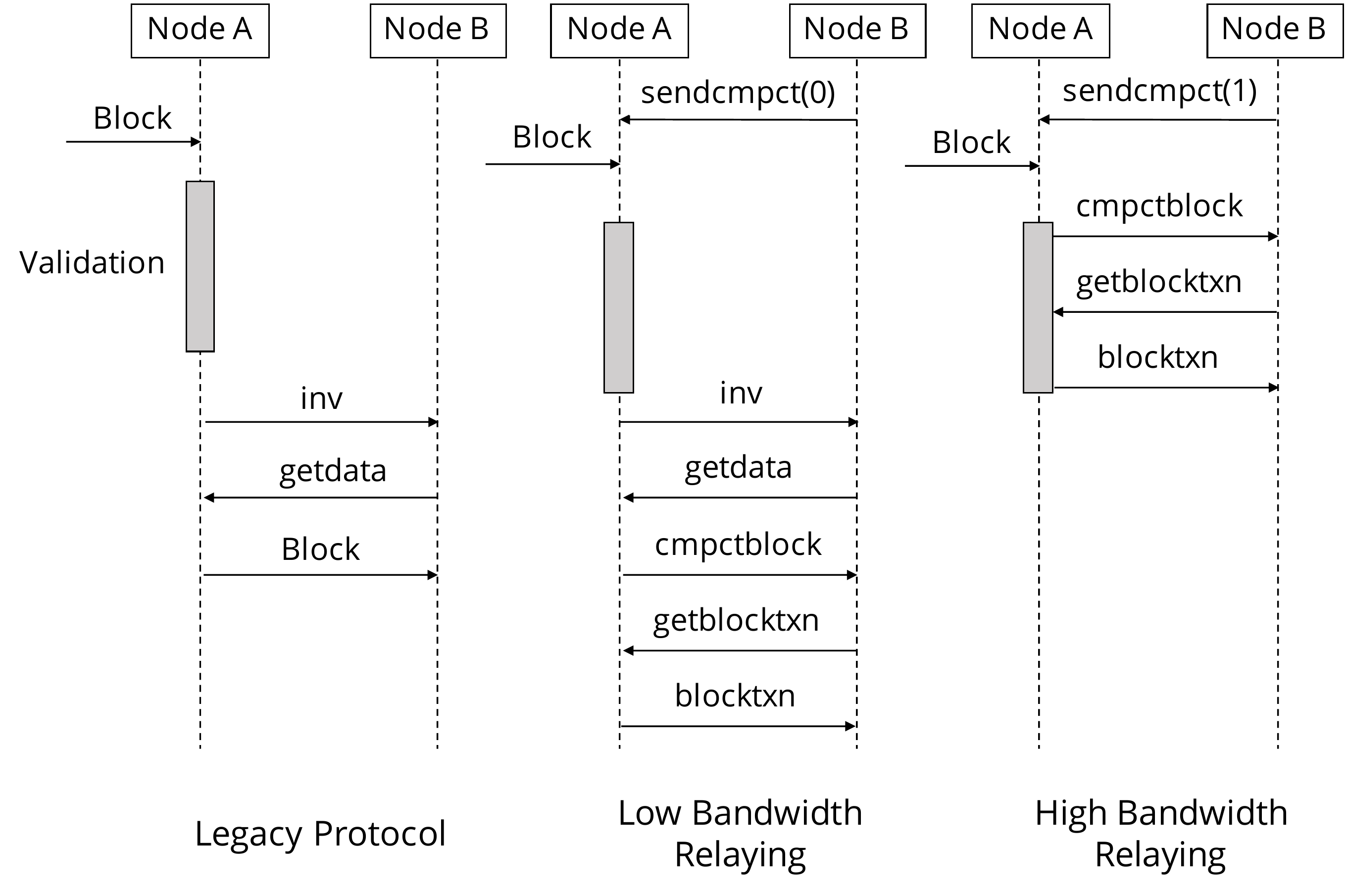}}
    \caption{Block propagation in legacy protocol and compact block relay.}
    \label{fig:compact-block-flow}
\end{figure}
Fig. \ref{fig:compact-block-flow} displays the block propagation flow in the legacy protocol and CBR.
In the legacy protocol, after receiving and verifying a new block, node A sends an \verb|inv| message containing the block's metadata to node B.
If node B has not received the block, it sends a \verb|getdata| message to node A requesting the entire block including the approved transactions.
In contrast, CBR only sends a compact block containing block headers and approved transaction indices.
If the receiving node fails to reconstruct a block from these data (i.e., if there is no transaction in the receiving node's memory pool), the receiving node requests the missing transactions from the sender node.

CBR has two optional protocols: low bandwidth relaying and high bandwidth relaying. 
In low bandwidth relaying, node A sends an \verb|inv| message.
In high bandwidth relaying, the confirmation that node B has already received the block is skipped, and node A sends the compact block as soon as the block is received before completing block validation.

\section{Simulating Bitcoin network}
We measured the block propagation delay using a simulator.
A simulator was used rather than an actual environment for the following three reasons:
\begin{itemize}
    \item The cost is lower than that of setting up nodes and building a network.
    \item The parameters, such as the number of nodes, network latency, and bandwidth can be easily changed.
    \item The impact of each factor on the block propagation delay can be distinguished and evaluated.
\end{itemize}

\begin{table}[t]
    \caption{parameter settings of bitcoin network.}
    \begin{center}
    \begin{tabular}{lll}
    \hline\hline
        Number of nodes & 6000 (2015) or 9000 (2019) \\
        Block generation interval & 10 min \\
        Block size & 535 KB (2015) or 1.0 MB (2019) \\
        Hash power & Gaussian distribution \\
        Number of neighbors & Distribution according to Miller et al.\cite{b10} \\
        Geographical distribution & Distribution according to Bitcoin \\
        \quad of nodes & \\
        Network latency & Six regional network latencies \\
        Network bandwidth & Six regional bandwidths \\
    \hline
    \end{tabular}
    \label{table:parameter-bitcoin-network}
    \end{center}
\end{table}

\begin{table}[t]
    \caption{sources of network parameters.}
    \begin{center}
    \begin{tabular}{lll}
    \hline\hline
         & 2015 & 2019 \\
         \hline
        Geographical distribution & Bitnodes\cite{b11} & Bitnodes\cite{b11} \\
        \quad of nodes & \\
        Network latency &  Verizon\cite{b16} & WonderNetwork\cite{b12}\\
        Network bandwidth & testmy.net\cite{b13} & testmy.net\cite{b13}\\
    \hline
    \end{tabular}
    \label{table:sources-of-network-parameters}
    \end{center}
\end{table}

In this study, we used the blockchain simulator SimBlock\cite{b7,b15}.
Because SimBlock can simulate block propagation between nodes, it can measure the block propagation delay.
Table \ref{table:parameter-bitcoin-network} presents the Bitcoin network parameters used in our simulation.

\subsection{Network parameters}
We used the network parameters presented in Aoki et al.\cite{b7} as network parameters for 2015 and calculated new network parameters for 2019.
Table \ref{table:sources-of-network-parameters} presents sources of network parameters for 2015 and 2019.
The network parameters included the node distribution, network latency, and bandwidth.
The calculation procedure of network parameters for 2019 is as follows.

\paragraph{Node distribution} We used data on the number of nodes in each country from Bitnodes\cite{b11}.
Because SimBlock has six regions (North America, Europe, South America, Asia, Japan, and Australia), the node distribution was calculated using the number of nodes in each region.

\paragraph{Network latency}
We selected a major city in the country (one city each in the east and west only for the United States), and obtained data on the network latency between the cities from WonderNetwork\cite{b12}.
The weighted average of the network latencies including the number of nodes in each country was used as the network latency between regions.

\paragraph{Bandwidth}
We used the bandwidth data of each country from testmy.net\cite{b13}.
The weighted average of the bandwidths, taking into account the number of nodes in each country, was used as the bandwidth of each region.

\subsection{Compact block relay model}
\begin{table}[t]
    \caption{parameter settings of compact block relay (CBR).}
    \begin{center}
    \begin{tabular}{lll}
    \hline\hline
    CBR Utilization Rate & 0.964 \\
    Compact Block Size & 18 KB \\
    Ratio of Churn Node & 0.976 \\
    Churn Node Block Reconstruction Failure Rate & 0.27 \\
    Control Node Block Reconstruction Failure Rate & 0.13 \\
    \hline
    \end{tabular}
    \label{table:parameter-cbr}
    \end{center}
    \end{table}
Table \ref{table:parameter-cbr} presents the parameters related to CBR.
We obtained the protocol version of each node from Bitnodes and adopted the percentage of nodes using the version of the protocol implementing CBR as the percentage of nodes using CBR.

Nodes using CBR should only behave as low bandwidth relaying.

\begin{figure}[t]
    \centerline{\includegraphics[width=\linewidth]{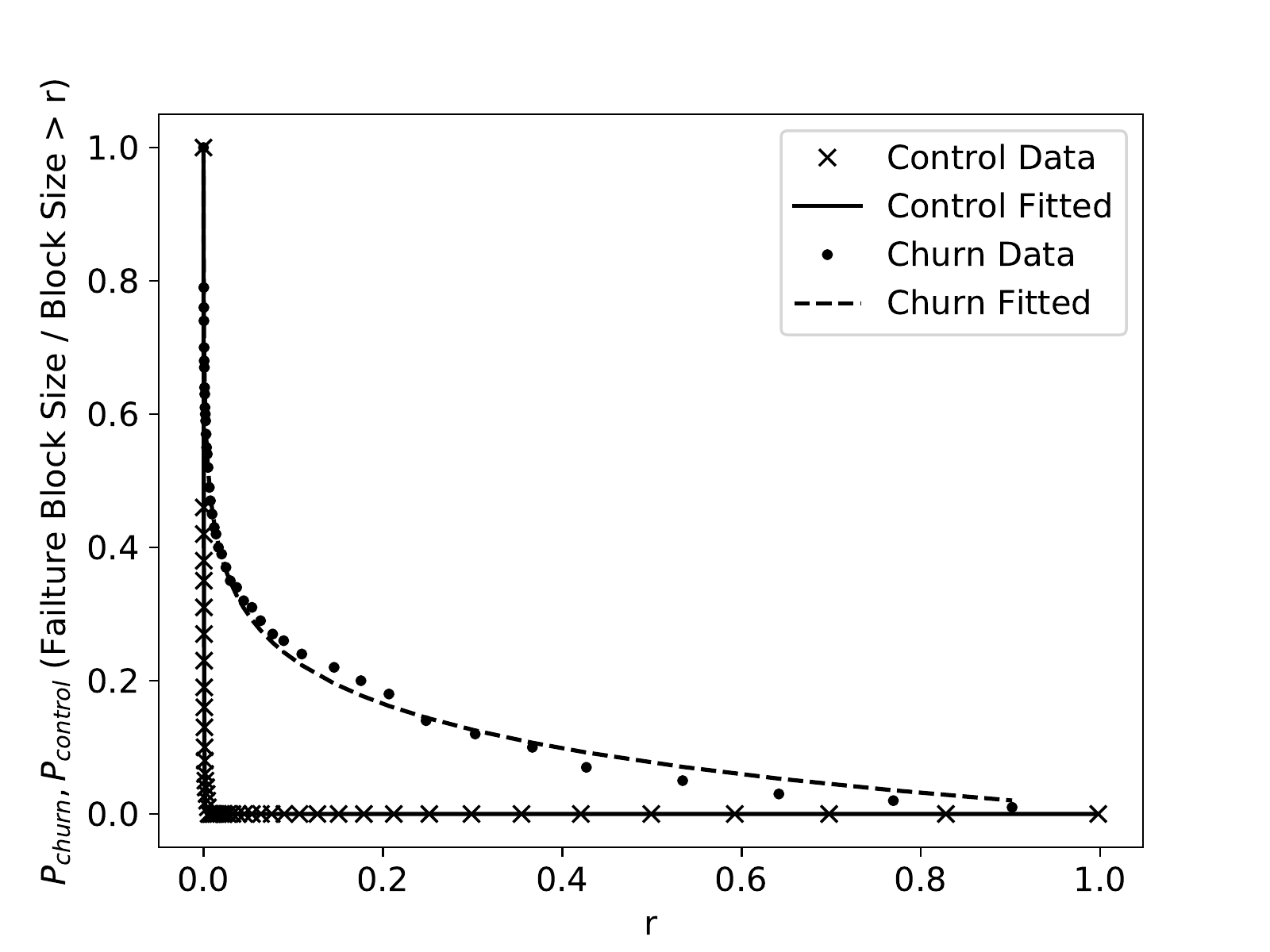}}
    \caption{Cumulative distribution of the failure block size.}
    \label{fig:block-size-distribution}
\end{figure}

According to Imtiaz et al.\cite{b14}, the failure rate of block reconstruction differs between the control node (the node that is always connected to the network) and the churn node (the node that repeats connection and disconnection).
The ratio of the control node and churn node and the failure rate of block reconstruction in low bandwidth relaying are based on the paper\cite{b14}.
Fig. \ref{fig:block-size-distribution} shows the cumulative distribution of the failure block size.
The failure block size, i.e., the data size to be downloaded upon reconstruction failure, follows the following cumulative distribution calculated from the number of missing transactions measured in the paper\cite{b14}.
$P_{churn}$ and $P_{control}$ represent the probability that the failure block size is larger than a given rate $r$ of the block size for churn node and control node.
\begin{align}
    P_{churn} &= e^{-2.12 \times 10^3 r} & (r \geq 0)\\
    P_{control} &= 1 - 0.0964\log( 2.89 \times 10^4 r + 1) & (r \geq 0)
\end{align}

\begin{table}[t]
    \caption{50th and 90th percentile of block propagation delay \protect\linebreak in real networks and a simulation.}
    \begin{center}
    \begin{tabular}{lrr}
    \hline\hline
         & 2015 Bitcoin network & 2019 Bitcoin network \\ 
    \hline
    50\% delay measured & 7,988 ms & 401 ms \\
    50\% delay simulated & 9,673 ms & 1,340 ms \\
    \hline
    90\% delay measured & 16,835 ms & 2,353 ms \\
    90\% delay simulated & 14,056 ms & 2,364 ms \\
    \hline
    \end{tabular}
    \label{table:compare-real-simulation}
    \end{center}
\end{table}

\section{Results and Discussion}
Here, we present the results of our evaluation.
\subsection{Simulator validation}

To experimentally validate our simulation, we compared the Bitcoin network in 2015 and 2019 with their respective simulated counterparts.
In the 2015 Bitcoin network simulation, the number of nodes was 6000, the block size was 535 KB, and the Internet parameters from 2015 were used.
In the simulation of the 2019 Bitcoin network simulation, the number of nodes was 9000, the block size was 1.0 MB, and CBR on Internet from 2019 was simulated.

Table \ref{table:compare-real-simulation} presents the simulation results and the measurement results of the block propagation delay obtained from a monitoring web site\cite{b2}.
The measured values are the monthly averages for July 2015 and October 2019.

Both 90th percentile simulated results were similar to the measurement results.
However, the 50th percentile results were larger than the measurement results.

This is likely due to the impact of the relay network. A relay network can perform block propagation efficiently to relay network participants using a relay server.
In our simulation, the network is a random network without a relay network; therefore, it takes several hops to reach 50\% of the network from the block generation nodes.
When using a relay network, the block generation node sends the block to the relay server and propagates it quickly to participating nodes, including nodes in other continents.
Thus, the measured propagation time to reach 50\% of the nodes is shorter than the simulated propagation time.
However, because nodes not participating in the relay network propagate blocks on a random network, if the participation rate in the relay network is not high, the 90th percentile of the propagation delay is not affected by the relay network as much as the 50th percentile.
The participation rate in Falcon, a relay network, was 2.65\% as of January 27, 2019~\cite{b4}.

Therefore, it is reasonable that the 50th percentile simulated results were larger than the measured results and the 90th percentile measured results were similar to the simulated results.
This means that our model correctly simulates Internet improvement and CBR.

\subsection{Block propagation delay}
\begin{figure}[t]
    \centerline{\includegraphics[width=\linewidth]{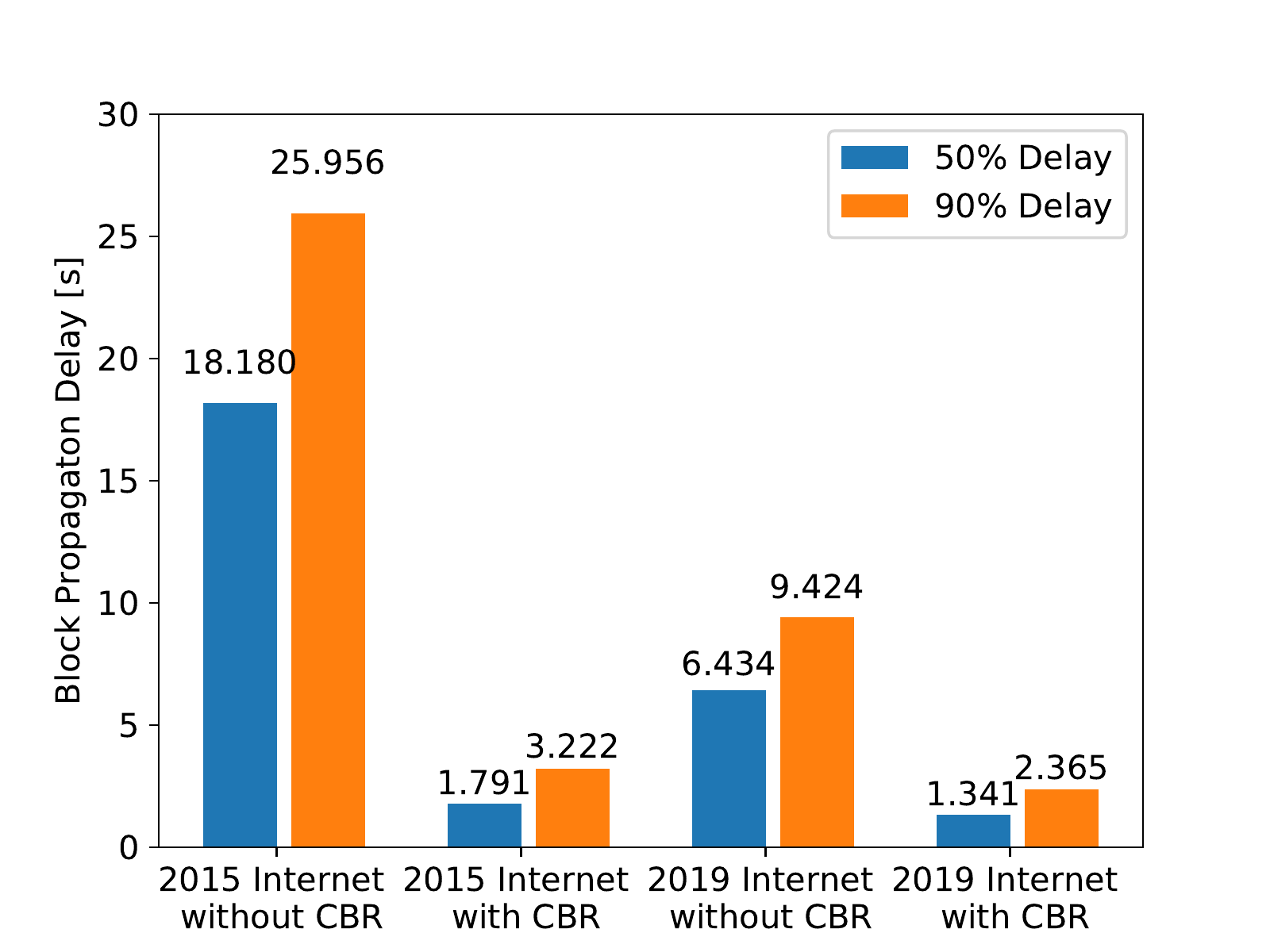}}
    \caption{50th and 90th percentile of block propagation delay.}
    \label{fig:propagation-delay-result}
\end{figure}

Here, we discuss the impact of Internet improvements and CBR on the block propagation delay.

In this experiment, simulations with and without CBR were performed with 2015 and 2019 Internet parameters.
The number of nodes was 9000, and the block size was 1.0 MB.

Fig. \ref{fig:propagation-delay-result} illustrates the 50th and 90th percentile of the block propagation delay measured.
By comparing the result of simulation with 2015 Internet without CBR and the result of simulation with 2015 Internet and CBR, the propagation delay is reduced by 90.1\% for the 50th percentile and 87.6\% for the 90th percentile.
Furthermore, comparing the result of simulation with 2015 Internet without CBR and the result of simulation with 2019 Internet without CBR, the propagation delay is reduced by 64.6\% for the 50th percentile and by 63.7\% for the 90th percentile.
These results reveal that CBR greatly reduces the propagation delay.

In terms of Internet improvements from 2015 to 2019, network latency became 0.889 times shorter on average, and bandwidth became 2--3 times wider.
In CBR, the compact block size was 0.018 times the block size of the legacy protocol.
The product of the data size and bandwidth accounts for most of the propagation delay; thus, CBR has a greater impact on the block propagation delay than Internet improvements.

Finally, the propagation delay was shorter in the simulation with 2019 Internet using CBR than in the simulation with 2019 Internet without CBR.
This is because the failure block size was still large; thus, the communication time was shortened by improving the bandwidth and the network latency was also shortened.

\subsection{Fork rate}
\begin{figure}[t]
    \centerline{\includegraphics[width=\linewidth]{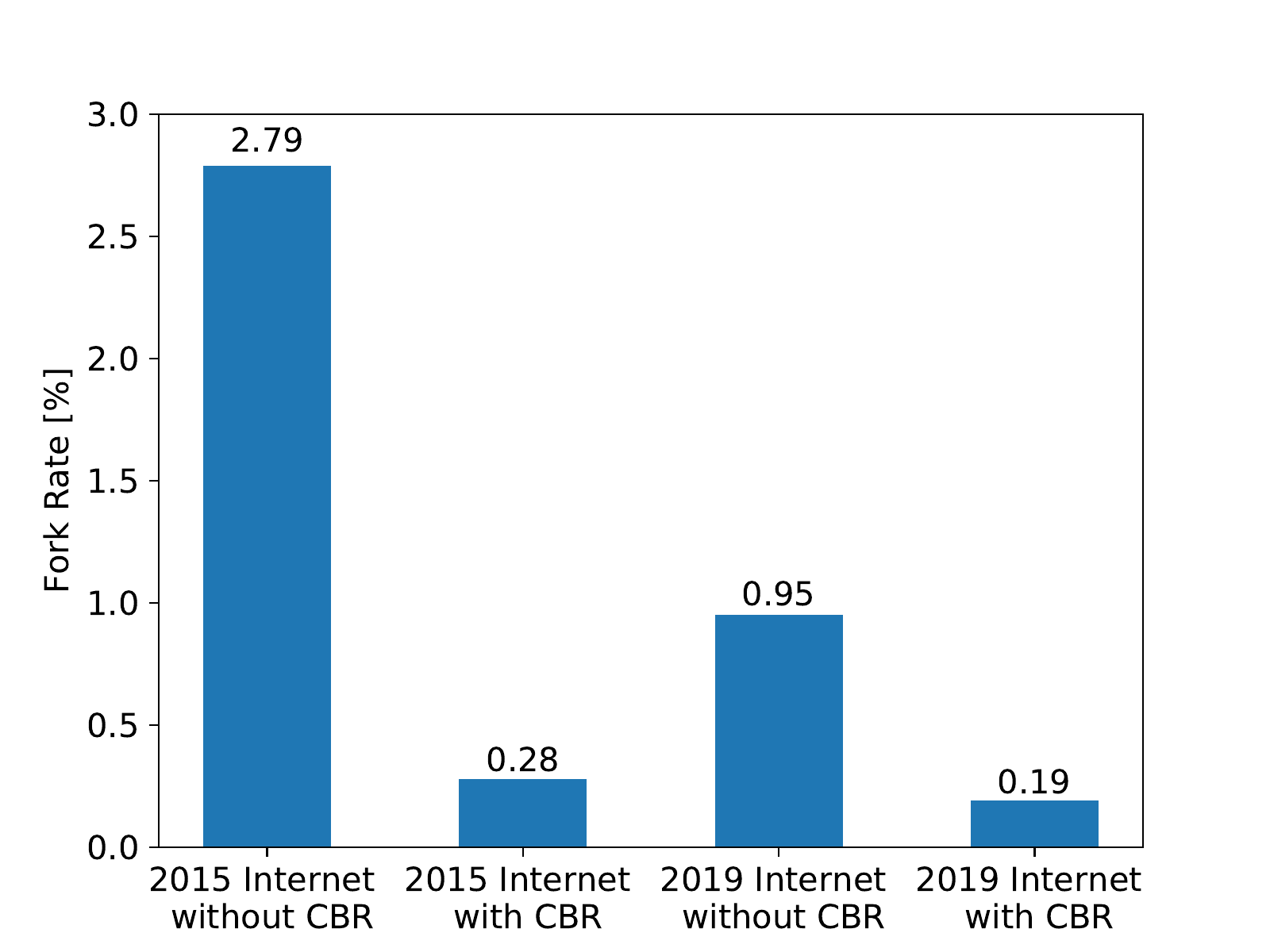}}
    \caption{Fork rate.}
    \label{fig:fork-rate-result}
\end{figure}

We also investigated the impact on fork rate. Fig. \ref{fig:fork-rate-result} shows the measured fork rates, which represent the ratio of orphan blocks among all generated blocks.
The fork rate is improved by shortening the block propagation delay; therefore, the security of the 2019 Bitcoin is superior that of the 2015 Bitcoin network.

\section{Conclusion}
In this study, we investigated the impacts of Internet improvements and CBR on the block propagation delay and fork rate by simulation.
The results indicate that CBR contributes to shortening the block propagation delay more than Internet improvements.
This is because the rate of block size reduction by CBR is larger than the rate of network latency reduction and bandwidth increase by  Internet improvements.

In addition, in the simulation of the Bitcoin network in 2015 and 2019, the 90th percentile of the block propagation delay was similar to the measurement results.
However, the value of the 50th percentile was larger than the measured value.
This is because the simulation assumed a random network; however, part of the actual Bitcoin network is a structured network such as a relay network.

Implementation of the topology and the node behavior of relay networks on SimBlock should improve the accuracy of the simulation.
This is left for future work and should make it possible to simulate a more realistic Bitcoin network.

\section*{Acknowledgment}
This work was supported by SECOM Science and Technology Foundation.


\begin{thebibliography}{00}
\bibitem{b1} Y. Sompolinsky and A. Zohar, “Secure high-rate transaction processing in bitcoin,” in International Conference on Financial Cryptography and Data Security. Springer, 2015, pp. 507--527.
\bibitem{b2} “Bitcoin Network Monitor - DSN Research Group, KASTEL @ KIT,” https://dsn.tm.kit.edu/bitcoin/, Accessed: Nov. 1. 2019.
\bibitem{b3} T. Neudecker, “Security and Anonymity Aspects of the Network Layer of Permissionless Blockchains,” Ph.D. thesis, Karlsruhe Institute of Technology (KIT), 2018.
\bibitem{b4} “Falcon - a fast bitcoin backbone,” https://www.falcon-net.org/, Accessed: Jan. 27. 2019.
\bibitem{b5} “Fibre fast internet bitcoin relay engine,” https://www.bitcoinfibre.org/, Accessed: Jan. 27. 2019.
\bibitem{b8} M. Corallo, “Compact Block Relay (BIP 152),”\\ https://github.com/bitcoin/bips/blob/master/bip-0152.mediawiki,\\ Accessed: Nov. 10. 2019.
\bibitem{b12} “Global Ping Statistics - WonderNetwork,” \\ https://wondernetwork.com/pings, Accessed: Oct. 20. 2019.
\bibitem{b13} “Top Countries for Bandwidth,” https://testmy.net/country, Accessed: Oct. 20. 2019.
\bibitem{b6} K. Otsuki, R. Banno, and K. Shudo, “Effects of a Simple Relay Network on the Bitcoin Network,” in Proc. Asian Internet Engineering Conference (AINTEC), August 2019.
\bibitem{b7} Y. Aoki, K. Otsuki, T. Kaneko, R. Banno, and K. Shudo, “SimBlock: A Blockchain Network Simulator,” in Proc. Workshop on Cryptocurrencies and Blockchains for Distributed Systems (CryBlock, In conjunction with IEEE INFOCOM 2019), April 2019.
\bibitem{b15} R. Banno, and K. Shudo, “Simulating a Blockchain Network with SimBlock,” in Proc. IEEE International Conference on Blockchain and Cryptocurrency (ICBC), May 2019.
\bibitem{b9} “Bitcoin Charts \& Graphs - Blockchain,”\\ https://www.blockchain.com/en/charts, Accessed: Oct. 10. 2019.
\bibitem{b10} A. Miller, J. Litton, A. Pachulski, N. Gupta, D. Levin, N. Spring, and B. Bhattacharjee, “Discovering bitcoins public topology and influential nodes,” 2015.
\bibitem{b11} “Bitnodes: Global Bitcoin Nodes Distribution,”\\ https://bitnodes.earn.com/, Accessed: Oct. 20. 2019.
\bibitem{b16} “Verizon latency,”\\ http://www.verizonenterprise.com/about/network/latency/, Not availble at Oct. 20. 2019.
\bibitem{b14} M. A. Imtiaz, D. Starobinski, A. Trachtenberg and N. Younis, “Churn in the Bitcoin Network: Characterization and Impact,” 2019 IEEE International Conference on Blockchain and Cryptocurrency (ICBC), 2019, pp. 431--439.
\end{thebibliography}
\end{document}